\documentclass[conference]{IEEEtran}
\IEEEoverridecommandlockouts
\usepackage{cite}
\usepackage{amsmath,amssymb,amsfonts}
\usepackage{graphicx}
\usepackage[ruled,vlined]{algorithm2e}
\usepackage{textcomp}
\usepackage[caption=false,font=footnotesize]{subfig}
\usepackage{xcolor}
\def\BibTeX{{\rm B\kern-.05em{\sc i\kern-.025em b}\kern-.08em
    T\kern-.1667em\lower.7ex\hbox{E}\kern-.125emX}}
\begin{document}

\title{Mean Spatial Frequency Decoupling for Learning-Based Uplink-to-Downlink Covariance Conversion in FDD Massive MIMO\\
}

\author{\IEEEauthorblockN{Melih Can Zerin}
\IEEEauthorblockA{\textit{Department of Electrical and Electronics Engineering} \\
\textit{Middle East Technical University}\\
Ankara, Türkiye \\
melih.zerin@metu.edu.tr}
\thanks{This work has been submitted to the IEEE for possible publication. Copyright may be transferred without notice, after which this version may no longer be accessible.}
}

\maketitle

\begin{abstract}
In frequency division duplexing (FDD) massive multiple-input multiple-output
(MIMO) systems, the uplink (UL)-to-downlink (DL) channel covariance matrix
(CCM) conversion problem is studied to relieve the heavy burden of DL training
and feedback required for channel estimation. Learning-based methods perform
well up to a certain array size, but for a fixed dataset size their accuracy
deteriorates with the number of antennas, to the point where simple
model-based methods outperform them. This paper identifies a key cause of this
behavior and addresses it. The mean angle of arrival (AoA) induces a phase ramp
along the lags of the CCM. Since the oscillation rate of this ramp grows with
the number of antennas, a dataset of fixed size becomes increasingly sparse
relative to the variation that must be captured. We propose estimating the
slope of this ramp from the UL CCM separately and mapping it to the DL band in
closed form, leaving the learner with a residual that is largely insensitive to
the mean AoA, which substantially reduces the performance degradation with an
increasing number of antennas. The proposed scheme, termed deramping, is a
combination of pre- and post-processing steps that applies to learning-based
conversion methods without altering their internal structure, as demonstrated
on three structurally different learners. Simulation results show that
deramping reduces the covariance estimation error of all three learners under
uniform, Laplacian, and Gaussian angular power spectra, keeps the
interpolation-based learners ahead of a model-based benchmark at large array
sizes, and improves downlink channel estimation.
\end{abstract}


\section{Introduction}
\label{sec1}

Massive multiple-input multiple-output (MIMO) is of vital importance for 5G and beyond networks owing to its high spectral efficiency and low energy consumption \cite{bjornson2017massive}.
Implementing it in frequency division duplexing (FDD) systems, however, is hampered by an excessive pilot and feedback overhead \cite{bjornson2016massive,xu2014user}, since the channel reciprocity of time division duplexing (TDD) systems does not hold for FDD systems \cite{zhong2020fdd} and the overhead grows with the number of base station antennas.
Exploiting second-order channel statistics such as the channel covariance matrix (CCM) helps reduce this burden \cite{xie2018channel}.
Several studies estimate the downlink (DL) CCM from the uplink (UL) CCM \cite{liang2001downlink,jordan2009conversion,decurninge2015channel,khalilsarai2018fdd,miretti2021channel,banerjee2022downlink,10189874,zerin2025downlink}, relying on the spatial reciprocity between the two bands \cite{hugl2002spatial}, which gives them the same power angular spectrum (PAS). 
Learning-based conversion methods \cite{decurninge2015channel,banerjee2022downlink,zerin2025downlink} can lose accuracy as the number of antennas grows for a fixed dataset size \cite{zerin2025downlink}.
Under a small angular spread, the covariance can be approximated by a
mean-angle phase term times an angular-spread term, which has been used for
parametric DL covariance estimation \cite{fang2017low}.
A related model-based scheme applies a diagonal phase transformation at the
dominant UL direction to the UL CCM \cite{liao2023reciprocity}, which adapts
the mean-angle phase term to the DL band but leaves the angular-spread term
unchanged.
In instantaneous DL channel extrapolation, the path delays and angles shared
by the two links have been extracted so that a neural network maps only the
frequency-dependent path gains \cite{choi2020downlink},and a path-oriented extrapolator that aligns each extracted path response has been proposed to improve generalization across environments
\cite{wang2026generalizable}.
Unlike these works, we remove the mean-angle phase term from both links
before learning and restore it analytically afterwards, so that a learner
maps only the angular-spread term. This requires neither resolvable paths nor
a particular PAS model and leaves the learner unchanged.

The main contributions of this paper are as follows.
\begin{itemize}
\item We identify a key cause of the performance degradation of learning-based
UL-to-DL CCM conversion with increasing array size. The mean spatial frequency,
i.e., the centroid of the PAS in the $\sin\phi$ domain,
enters the CCM as a phase ramp whose phase at the $m$th lag sweeps an interval
of width $2\pi m$, so a fixed-size training set covers the data ever more
sparsely as the number of antennas $M$ grows. In terms of the error bound of
\cite[Theorem~2]{zerin2025downlink}, this can be interpreted as a growth of
$\delta$, the distance from a test point to its nearest training neighbor, and
removing the ramp before learning is expected to slow this growth, which is
consistent with the reduced degradation observed in Section~\ref{Simulation_Section}.
\item We propose deramping, which removes the ramp from both links before
learning and restores it analytically afterwards, since its slope scales
exactly by the carrier frequency ratio $f_R$ \cite{liao2023reciprocity}. The deramp and restore
operations are exactly invertible for any estimate of the mean spatial
frequency, so an estimation error only leaves a small residual slope for the
learner, and setting the estimate to zero recovers the baseline method
identically. The learner itself is left unchanged, and we demonstrate the
scheme on three structurally different learners.
\item It was shown in \cite{zerin2025downlink} that learning-based methods are
outperformed by the sinc transformation \cite{10189874} as $M$ grows. With
deramping, the dictionary \cite{decurninge2015channel} and kernel interpolators \cite{zerin2025downlink} remain ahead of the sinc
transformation at all tested array sizes, and the gains carry over to DL
channel estimation and to non-uniform PAS models.
\end{itemize}

\textit{Notation.} Bold lower-case and upper-case letters denote vectors and matrices, respectively. $(\cdot)^T$, $(\cdot)^H$, $\mathbf{A}^{-1}$, and $\mathrm{tr}(\mathbf{A})$ denote the transpose, Hermitian transpose, inverse, and trace, respectively.

\section{System Model} \label{System_Model_Section}
The system model of \cite{zerin2025downlink} is adopted, where the base station (BS) has a uniform linear array (ULA) of $M$ antennas serving single-antenna user equipment (UE). The carrier frequency and wavelength of the UL band are $f_{\mathrm{UL}}$ and $\lambda_{\mathrm{UL}}$, and those of the DL band are $f_{\mathrm{DL}}$ and $\lambda_{\mathrm{DL}}$. Their ratio is $f_R=f_{\mathrm{DL}}/f_{\mathrm{UL}}$, and the antenna spacing is $d=\lambda_{\mathrm{UL}}/2$.

The wide sense stationary uncorrelated scattering (WSSUS) model is adopted as in \cite{zerin2025downlink}.
The CCMs of the zero-mean, frequency-flat UL and DL channels are \cite{zerin2025downlink}

\begin{multline} \label{CCM_Formula}
    \mathbf{R}^{x} = 
    \mathop{\mathbb{E}} \left\{\mathbf{h}^{x} (\mathbf{h}^{x})^{H} \right\} =  \int_{\bar{v}-\Delta}^{\bar{v}+\Delta} p(\phi)\mathbf{a}_{x} (\phi) \mathbf{a}_{x}^{H} (\phi) \,d\phi, \\ x \in \{\mathrm{UL},\mathrm{DL}\},
\end{multline}
where $\mathbf{h}^{x}$ is the channel vector, $p(\phi)$ is the PAS, $\Delta$ is its half-width, $\bar{v}$ is the mean angle of arrival (AoA), and $\mathbf{a}_x(\phi)=[1,\,e^{j\pi\gamma_x\sin\phi},\,\dots,\,e^{j\pi\gamma_x(M-1)\sin\phi}]^T$ is the array response vector with $\gamma_{\mathrm{UL}}=1$ and $\gamma_{\mathrm{DL}}=f_R$. 

Each CCM is Hermitian and Toeplitz, and hence determined by its first column, whose $m$th entry is

\begin{equation} \label{r_ccm_eqn_phi_domain}
    r^{x}_{m} = \int_{\bar{v}-\Delta}^{\bar{v}+\Delta} p(\phi)\, e^{j\pi\gamma_{x} m \sin\phi} \,d\phi, \quad m=0,\dots,M-1.
\end{equation}

Defining the spatial frequency $u=\sin\phi$, \eqref{r_ccm_eqn_phi_domain} becomes

\begin{equation} \label{r_ccm_eqn_u_domain}
    r^{x}_{m} = \int_{-1}^{1} \rho(u)\, e^{j\pi\gamma_{x} m u} \,du,
\end{equation}

where $\rho(u)$ is the PAS in the $u$-domain.

\section{Downlink Covariance Estimation via Deramping} \label{DL_CCM_Estimation_Section}

\subsection{Problem Formulation}

Given $N$ training pairs $\{(\mathbf{r}^{\mathrm{UL}}_i,\mathbf{r}^{\mathrm{DL}}_i)\}_{i=1}^{N}$,
learning-based methods seek a mapping
$f:\mathbf{r}^{\mathrm{UL}}\mapsto\mathbf{r}^{\mathrm{DL}}$ that generalizes to
unseen UL CCMs. For a ULA with $d=\lambda_{\mathrm{UL}}/2$, the covariance lags
can be factorized exactly as
\begin{equation}
  r^{x}_m = e^{\,j\pi\gamma_x m u_0}\, c^{x}_m.
  \label{eq:factorization}
\end{equation}

The first factor $e^{\,j\pi\gamma_x m u_0}$ is linear in phase with respect to the lag index $m$, and we
refer to it as the \emph{phase ramp}. Its slope $\pi\gamma_x u_0$ is set by the
\emph{mean spatial frequency} $u_0$, defined as the centroid of the PAS in the
domain $u=\sin\phi$. The second factor $c^{x}_m$, which we call the \emph{residual},
depends only on the shape of the PAS around $u_0$.
Since the PAS is reciprocal between the two links, the same $u_0$ governs both, and only the slope of the ramp scales with $f_R$.
The phase ramp dominates the variability of the CCM entries. As $u_0$ spans
$[-1,1]$, the ramp phase of the $m$th lag sweeps an interval of width $2\pi m$. A
fixed training set therefore covers the data ever more sparsely as $M$ grows.
The two factors transform differently from UL to DL. The ramp phase scales
exactly, $\pi f_R m u_0 = f_R(\pi m u_0)$, whereas the residual does not,
because $c^{\mathrm{DL}}_m$ is the residual of the same PAS evaluated at the
non-integer lag $f_R m$. A direct mapping must therefore handle two operations
implicitly. For the ramp, it must scale an unwrapped phase by the non-integer
factor $f_R$, although each entry reveals only the wrapped sum of the ramp and
residual phases. The number of $2\pi$ wraps must thus be inferred from the
entire vector. For the residual, it must interpolate a slowly varying function
to non-integer lags, which is the part that actually requires learning. Since the
residual is also largely insensitive to $u_0$, this motivates estimating $u_0$ from the UL CCM, removing the ramp
from both links before learning, and restoring it analytically afterwards.

\subsection{Estimation of the Mean Spatial Frequency} \label{subsec_mean_spat_est}

We estimate $u_0$ from the Bartlett
spectrum
$B(u) = \mathbf{a}^H(u)\,\mathbf{R}^{\mathrm{UL}}\mathbf{a}(u)$ \cite{van2002optimum}, where $\mathbf{a}(u)$ denotes $\mathbf{a}_{\mathrm{UL}}$ expressed in the $u$-domain.
Owing to the Toeplitz structure of the CCM, it can be computed directly
from the UL lags as
\begin{equation} \label{eq:bartlett}
  B(u) = \sum_{k=-(M-1)}^{M-1} (M-|k|)\, r^{\mathrm{UL}}_k\, e^{-j\pi k u}.
\end{equation}
$B(u)$ is not the PAS itself but its convolution with the power pattern of
the uniformly weighted array, $F_M(x) = \sin^2(\pi M x/2)/\sin^2(\pi x/2)$, whose main lobe has
width $4/M$ and whose sidelobes decay only as $1/x^2$ over the entire
periodic $u$-axis.
The mean spatial frequency is the centroid of the PAS in the $u$-domain.
Computing the same centroid from $B(u)$ over a full period, however, is
unreliable. Since the $u$-axis is periodic, the linear centroid depends on
where the period is cut and is dominated by the slowly decaying sidelobes of
$F_M$. The circular centroid, in turn, reduces to the phase of the lag-1
entry, and hence uses only the correlation between adjacent antennas while
discarding the rest of the aperture. We therefore restrict the centroid
computation to the main-lobe region of $B(u)$:
\begin{enumerate}
  \item $B(u)$ is evaluated on a uniform grid of $N_u = 16M$ points over
  $[-1,1)$ via a zero-padded discrete Fourier transform, which improves the
  numerical accuracy of the peak and centroid but not the resolution.
  \item Let $u_{\mathrm{pk}}$ denote the peak of $B(u)$, and let
  $\mathcal{S}$ be the connected region around it on which
  $B(u) > \alpha \max_u B(u)$. We set $\alpha = 0.1$ ($-10$~dB), just
  above the first sidelobe of $F_M$ ($\approx -13$~dB), so that
  $\mathcal{S}$ covers the main lobe but not the sidelobes. The region is
  taken circularly and may wrap around $u = \pm 1$.
  \item The estimate is the centroid of $B$ over $\mathcal{S}$, computed
  relative to the peak:
  \begin{equation} \label{eq:u0hat}
    \hat{u}_0 = u_{\mathrm{pk}} + \frac{\sum_{u\in\mathcal{S}}
      B(u)\,(u - u_{\mathrm{pk}})}{\sum_{u\in\mathcal{S}} B(u)},
  \end{equation}
  where $u - u_{\mathrm{pk}}$ is measured along the circle.
\end{enumerate}
No phase unwrapping is involved, since $[-1,1)$ is exactly one period of
$B(u)$ for $d = \lambda_{\mathrm{UL}}/2$, and the estimator does not assume
any particular PAS shape.

\subsection{Proposed Method} \label{sec:proposed}

The proposed method wraps a learning-based UL-to-DL conversion method with a
pre-processing and a post-processing stage, as summarized in
Algorithm~\ref{alg:deramp}. The operation $\textsc{EstimateMSF}(\cdot)$
returns the estimate $\hat u_0$ of the mean spatial frequency (MSF) of a UL
CCM (Section~\ref{subsec_mean_spat_est}), and $\textsc{Deramp}$ removes a
phase ramp from a lag sequence:
\begin{equation} \label{eq:deramp}
  [\textsc{Deramp}(\mathbf{r}, u, \gamma)]_m = r_m\, e^{-j\pi\gamma m u},
  \qquad m = 0,\dots,M-1.
\end{equation}
We denote by $\tilde{\mathbf{c}}$ the lags deramped with an estimate $\hat u_0$, which coincide with the residual in \eqref{eq:factorization} when $\hat u_0=u_0$.

In the training stage, $\hat u_{0,i}$ is estimated from the UL CCM of each
training sample, and both the UL input and the DL target of that sample are
deramped with this same estimate, the DL slope being scaled by $f_R$ as in
\eqref{eq:factorization}. Using the UL estimate for the DL target is essential, because only the UL CCM is available at test time, and it keeps deramping and ramp restoration consistent. The deramped lags are stacked as
real vectors in the same format as the baseline, so the learner is trained
without modification.

In the test stage, the UL CCM is deramped with its own estimate
$\hat u_{0,\ast}$, the learner predicts the DL residual, and the DL ramp
$e^{j\pi f_R m \hat u_{0,\ast}}$ is restored analytically before the DL CCM
is assembled as the Hermitian Toeplitz matrix with first column $\hat{\mathbf{r}}^{\mathrm{DL}}_\ast$.

\begin{algorithm}[!t]
\SetKwInOut{Input}{Input}\SetKwInOut{Output}{Output}
\SetKwFor{ForEach}{for each}{do}{end}
\SetAlgoVlined
\Input{Training pairs $\{(\mathbf{r}^{\mathrm{UL}}_i,\mathbf{r}^{\mathrm{DL}}_i)\}_{i=1}^{N}$,
       test input $\mathbf{r}^{\mathrm{UL}}_\ast$, frequency ratio $f_R$}
\Output{Estimated DL CCM $\hat{\mathbf{R}}^{\mathrm{DL}}_\ast$}
\BlankLine
\textbf{Phase 1: deramp both links with the UL estimate}\;
\ForEach{$i=1,\dots,N$}{
  $\hat{u}_{0,i} \leftarrow \textsc{EstimateMSF}(\mathbf{r}^{\mathrm{UL}}_i)$\;
  $\tilde{\mathbf{c}}^{\mathrm{UL}}_i \leftarrow
     \textsc{Deramp}(\mathbf{r}^{\mathrm{UL}}_i,\hat{u}_{0,i},1)$\;
  $\tilde{\mathbf{c}}^{\mathrm{DL}}_i \leftarrow
     \textsc{Deramp}(\mathbf{r}^{\mathrm{DL}}_i,\hat{u}_{0,i},f_R)$\;
}
\BlankLine
\textbf{Phase 2: train the learner (left unchanged)}\;
$f(\cdot) \leftarrow \textsc{TrainLearner}\big(\{(\tilde{\mathbf{c}}^{\mathrm{UL}}_i,
   \tilde{\mathbf{c}}^{\mathrm{DL}}_i)\}_{i=1}^{N}\big)$\;
\BlankLine
\textbf{Phase 3: deramp, predict, and restore the ramp}\;
$\hat{u}_{0,\ast} \leftarrow \textsc{EstimateMSF}(\mathbf{r}^{\mathrm{UL}}_\ast)$\;
$\tilde{\mathbf{c}}^{\mathrm{UL}}_\ast \leftarrow
   \textsc{Deramp}(\mathbf{r}^{\mathrm{UL}}_\ast,\hat{u}_{0,\ast},1)$\;
$\hat{\mathbf{c}}^{\mathrm{DL}}_\ast \leftarrow f(\tilde{\mathbf{c}}^{\mathrm{UL}}_\ast)$\;
normalize $\hat{\mathbf{c}}^{\mathrm{DL}}_\ast$ so that its first entry is unity\;
$\hat{r}^{\mathrm{DL}}_{\ast,m} \leftarrow \hat{c}^{\mathrm{DL}}_{\ast,m}\,
   e^{\,j\pi f_R m \hat{u}_{0,\ast}}$,\quad $m=0,\dots,M-1$\;
$\hat{\mathbf{R}}^{\mathrm{DL}}_\ast \leftarrow
   \mathrm{Toeplitz}(\hat{\mathbf{r}}^{\mathrm{DL}}_\ast)$\;
\caption{DL CCM estimation via deramping}
\label{alg:deramp}
\end{algorithm}

\section{Simulation Results} \label{Simulation_Section}

\begin{figure*}[!t]
\centering
\subfloat[]{\includegraphics[width=2.1in]{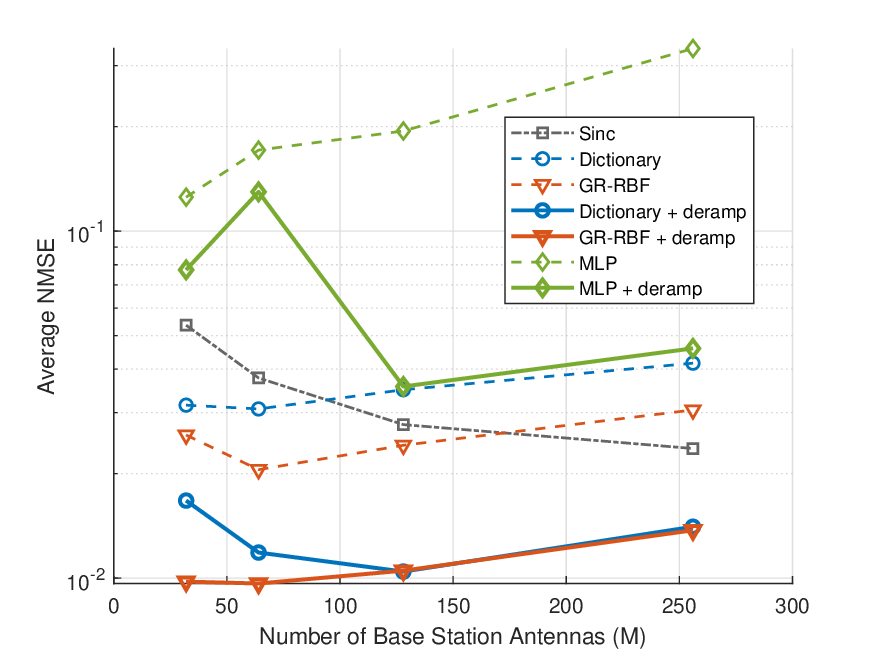}\label{M_vs_NMSE}}\hfil
\subfloat[]{\includegraphics[width=2.1in]{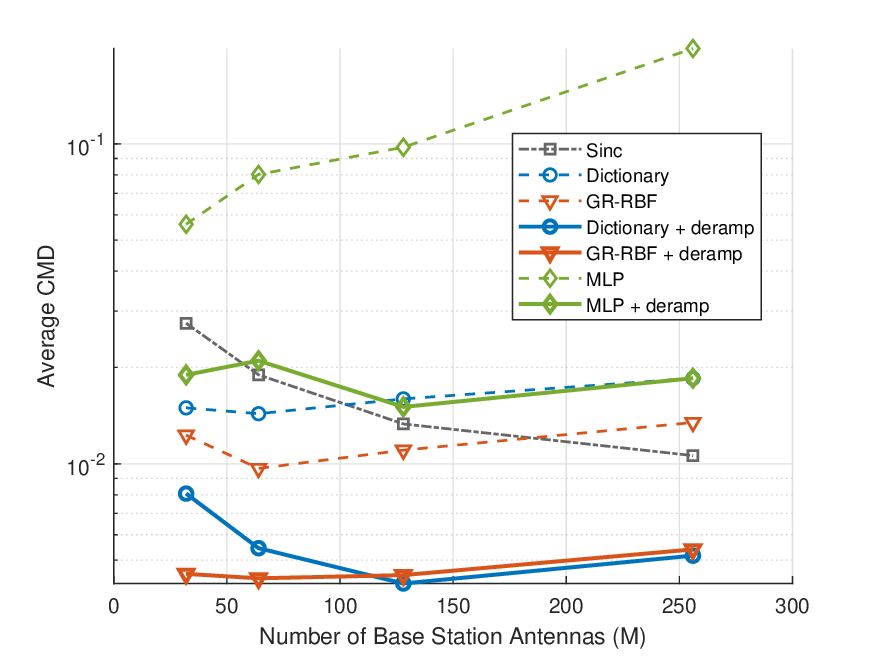}\label{M_vs_CMD}}\hfil
\subfloat[]{\includegraphics[width=2.1in]{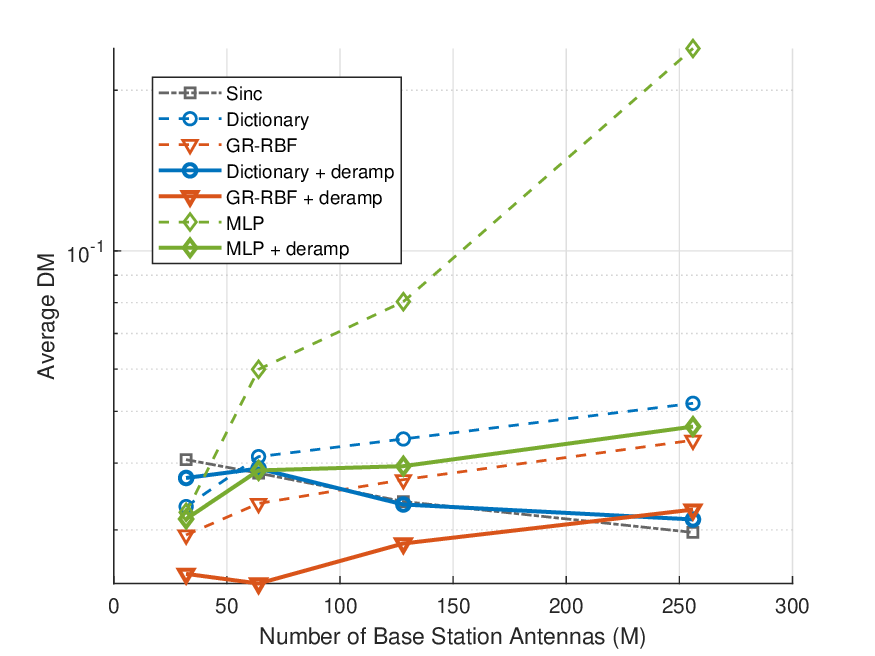}\label{M_vs_DM}}
\caption{Errors of the compared methods versus the number of base station
antennas at an SNR of $20$~dB: (a)~NMSE, (b)~CMD, (c)~DM. The spike of the
deramped MLP at $M=64$ in (a) is caused by a single failed training out
of $25$.}
\label{M_vs_Error}
\end{figure*}

\begin{figure*}[!t]
\centering
\subfloat[]{\includegraphics[width=2.1in]{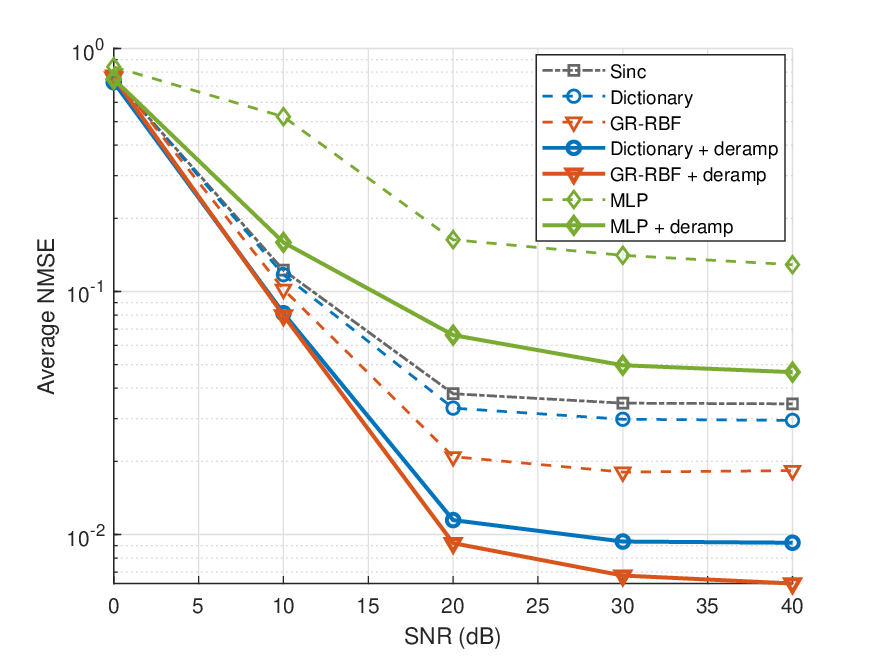}\label{SNR_vs_NMSE}}\hfil
\subfloat[]{\includegraphics[width=2.1in]{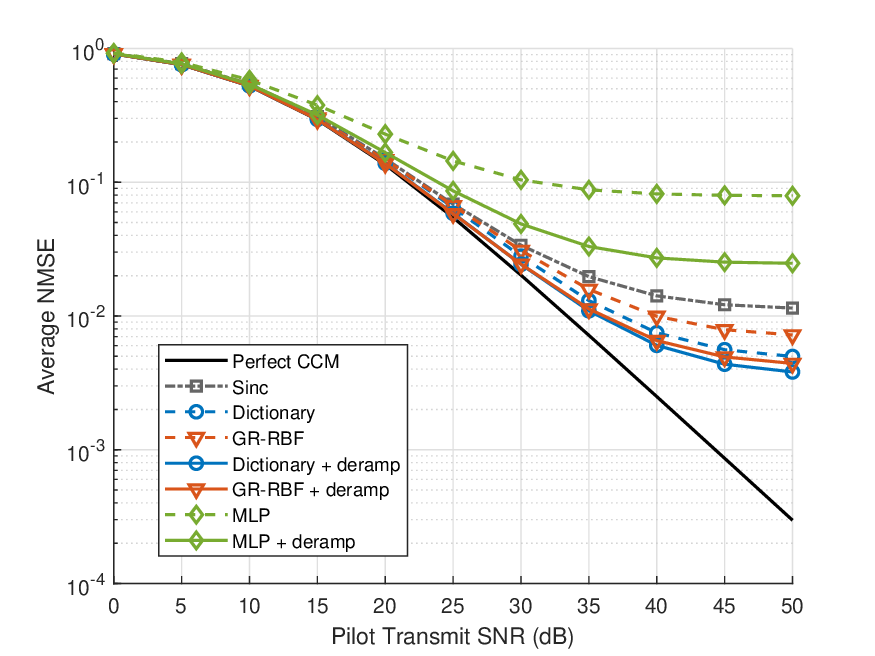}\label{NMSE_vs_SNR_for_MMSE_ch_est}}\hfil
\subfloat[]{\includegraphics[width=2.1in]{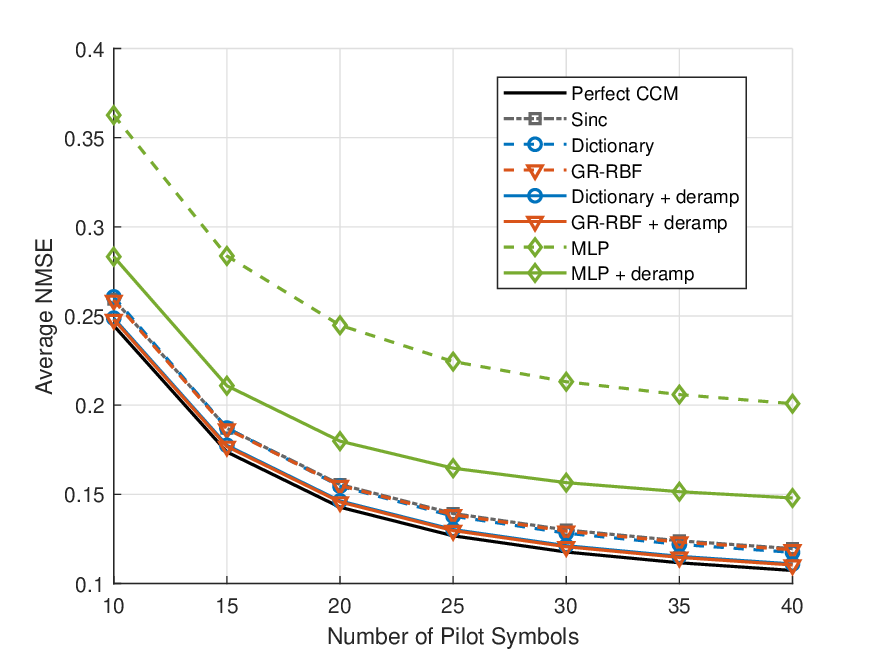}\label{NMSE_vs_T_for_MMSE_ch_est}}
\caption{(a)~NMSE of the CCM estimates versus SNR for $M=64$. NMSE of the
MMSE channel estimator versus (b)~the pilot SNR and (c)~the number of pilot
symbols.}
\label{SNR_vs_Error}
\end{figure*}

The experiments are conducted by varying $M$ at a fixed signal-to-noise ratio
(SNR) and by varying the SNR at a fixed $M$. Unless otherwise stated, $M = 64$, the SNR is $20$~dB, and
the results are averaged over $25$ independent and identically distributed
(i.i.d.) datasets. The simulation parameters are listed in
Table~\ref{sim_param_table}.

\begin{table}[!t]
\caption{Simulation Parameters}
\label{sim_param_table}
\centering
\begin{tabular}{|p{3.2cm}|p{4.3cm}|}
 \hline
 Carrier frequencies & $f_{\mathrm{UL}} = 1.95$~GHz, $f_{\mathrm{DL}} = 2.14$~GHz \\
 \hline
 Antenna spacing & $d = \lambda_{\mathrm{UL}}/2$ \\
 \hline
 Number of antennas $M$ & $\{32, 64, 128, 256\}$ \\
 \hline
 SNR & $\{0, 10, 20, 30, 40\}$~dB \\
 \hline
 PAS & Uniform \\
 \hline
 Mean AoA & $\sim\mathcal{U}[-\pi, \pi]$ \\
 \hline
 PAS half-width & $\sim\mathcal{U}[5^\circ, 15^\circ]$ \\
 \hline
 Channel realizations per CCM & $2M$ \\
 \hline
 Dataset size & $N_{\mathrm{train}} = 400$, $N_{\mathrm{test}} = 100$ \\
 \hline
 GR-RBF \cite{zerin2025downlink} &
   $\mu_1 = 0.1$, $\mu_2 = 3\times10^5$, $\mu_3 = 100$ \\
 \hline
 MLP hidden layers & $(256, 64, 256)$, ReLU \\
 \hline
 Deramping & $\alpha = 0.1$, $N_u = 16M$ \\
 \hline
\end{tabular}
\end{table}

The training and test datasets are generated as in \cite{zerin2025downlink}:
each CCM is a sample covariance computed from $2M$ channel realizations
observed at the given SNR, projected onto the set of Hermitian positive
semidefinite Toeplitz matrices using the method of
\cite{grigoriadis1994application}.

We apply deramping to three learning-based methods: the dictionary-based
interpolator of \cite{decurninge2015channel}, the graph-regularized RBF kernel
interpolator (GR-RBF) of \cite{zerin2025downlink}
and a small
multilayer perceptron (MLP) trained with Adam and early stopping.
The MLP is not intended as a competitive method. It is
included to test if the gains extend to neural networks. Each learner is
evaluated with and without deramping under identical settings. As a
model-based reference, we include the sinc transformation \cite{10189874},
which requires no training and was shown in \cite{zerin2025downlink} to
outperform learning-based methods at large $M$.

Performance is measured by the three metrics of \cite{zerin2025downlink}:
normalized mean squared error (NMSE), correlation matrix distance (CMD), and
deviation metric (DM). Fig.~\ref{M_vs_Error} shows the metrics as a function
of $M$ at $20$~dB, and Fig.~\ref{SNR_vs_Error}(a) shows the NMSE as a function of the SNR at
$M = 64$. The CMD and DM versus SNR are omitted for brevity. Three
observations follow.

  \begin{itemize}
    \item \emph{Deramping improves all three learners.} At $M = 256$, it
  reduces the NMSE of the dictionary and GR-RBF methods by factors of
  $2.95$ and $2.23$, respectively. For the MLP, whose mean is dominated by
  occasional training failures discussed below, we compare median values.
  The median NMSE of the MLP is reduced by a factor that grows
  monotonically with $M$, from $2.5\times$ at $M = 32$ to $6.1\times$ at $M = 256$.
  This is consistent with the phase ramp becoming the dominant source of
  variability as $M$ grows.
    \item \emph{The crossover with the sinc transformation is removed.}
  Without deramping, the dictionary and GR-RBF methods fall behind the sinc
  transformation for $M \geq 128$ and $M = 256$, respectively. With
  deramping, both remain below it at every tested $M$, by a factor of about
  $1.7$ at $M = 256$. Their errors still increase from $M = 128$ to
  $M = 256$, but the absolute increase is roughly halved, from $0.0064$ to
  $0.0032$ for GR-RBF and from $0.0067$ to $0.0037$ for the
  dictionary method. The MLP remains above the sinc transformation at every
  $M$ even with deramping, which is consistent with its role as a probe
  rather than a competitive method.
  \item \emph{The gain persists across SNR values and increases with the SNR.}
  At low SNR, the error is dominated by the noise in the sample covariances
  rather than by the phase ramp, which limits the benefit of deramping.
\end{itemize}

\begin{table}[!t]
\caption{NMSE at $M=256$ Under Non-Uniform PAS Models}
\label{tab:nonuniform}
\centering
\begin{tabular}{l|cc|cc}
\hline
 & \multicolumn{2}{c|}{Laplacian} & \multicolumn{2}{c}{Gaussian} \\
Method & Raw & Deramped & Raw & Deramped \\
\hline
Dictionary \cite{decurninge2015channel} & 0.0199 & 0.0090 & 0.0174 & 0.0058 \\
GR-RBF \cite{zerin2025downlink} & 0.0147 & 0.0089 & 0.0083 & 0.0048 \\
MLP & 0.3598 & 0.0401 & 0.1614 & 0.0129 \\
\hline
Sinc \cite{10189874} & \multicolumn{2}{c|}{0.0223} & \multicolumn{2}{c}{0.0198} \\
\hline
\end{tabular}
\end{table}

The occasional failures of the deramped MLP stem from its per-feature
standardization, which amplifies the noise in the nearly constant imaginary
parts of the first residual lags. The interpolators do not standardize
features and are unaffected.

To assess the dependence on the PAS model, Table~\ref{tab:nonuniform}
reports the NMSE at $M=256$ for the Laplacian and Gaussian PAS models of
\cite{zerin2025downlink}, averaged over $10$ datasets. Deramping improves
all three learners under both models. The gains are $1.6$--$1.7\times$ for
GR-RBF, $2.2$--$3.0\times$ for the dictionary method, and
about an order of magnitude for the MLP, in the same order as under the
uniform PAS. This is expected, since the factorization
\eqref{eq:factorization} holds for any PAS and the estimator
\eqref{eq:u0hat} does not assume a particular PAS shape. Unlike the uniform
case, the dictionary and GR-RBF methods already outperform the sinc transformation
at $M=256$ under these models, and deramping widens the margin further.
Deramping also narrows the gap between the dictionary and GR-RBF methods,
from $35\%$ to $1\%$ under the Laplacian PAS and from $110\%$ to $21\%$
under the Gaussian PAS.

Since the DL CCM is ultimately used for DL channel estimation, we also
evaluate the minimum mean squared error (MMSE) channel state information (CSI)
estimation performance obtained with each DL CCM estimate.
The DL CCMs are estimated at $M=64$ and
$20$~dB SNR. After pilot transmission for DL CSI acquisition, the received
signal at the user equipment (UE) is
\begin{equation}
    \mathbf{y}=\mathbf{X}\mathbf{h}+\mathbf{n},
\end{equation}
where $\mathbf{h}\in\mathbb{C}^{M}$ is the DL channel vector with covariance
$\mathbf{R}^{\mathrm{DL}}$, $\mathbf{y}=[y_{1} \ \dots \ y_{N_{p}}]^{T}$
contains the $N_p$ received symbols, the rows of
$\mathbf{X}=[\mathbf{x}_{1} \ \dots \ \mathbf{x}_{N_{p}}]^{T}$ are the pilot
vectors, and $\mathbf{n}\sim\mathcal{CN}(\mathbf{0},\sigma_{p}^{2}\mathbf{I})$
is the noise vector. The MMSE DL CSI estimator is \cite{fang2017low}
\begin{equation} \label{MMSE_ch_est_formula}
    \hat{\mathbf{h}}_{\mathrm{MMSE}}=\mathbf{R}^{\mathrm{DL}}\mathbf{X}^{H}
    \left( \mathbf{X}\mathbf{R}^{\mathrm{DL}}\mathbf{X}^{H}
    +\sigma_{p}^{2}\mathbf{I} \right)^{-1}\mathbf{y}.
\end{equation}
In the experiments, $\mathbf{R}^{\mathrm{DL}}$ in \eqref{MMSE_ch_est_formula}
is replaced by the estimate obtained with each method.
The rows of the pilot
matrix are chosen to be orthogonal and scaled so that
$\mathbf{X}\mathbf{X}^{H}=(P/N_p)\mathbf{I}$, which satisfies the total pilot
power constraint $\mathrm{tr}(\mathbf{X}\mathbf{X}^{H})\le P$ with equality and
requires $N_p\le M$. The pilot SNR is defined as $P/\sigma_p^2$, and the
estimation performance is measured by the average of
$\|\hat{\mathbf{h}}-\mathbf{h}\|^2/\|\mathbf{h}\|^2$.

Fig.~\ref{SNR_vs_Error}(b) and~(c) show the performance for varying pilot SNR
and number of pilot symbols $N_p$, together with a genie-aided benchmark that uses the true
DL CCM. For each learning-based method, deramping improves the channel
estimation performance, so the gain in DL CCM estimation carries over to the
application for which the DL CCM is ultimately needed.

\section{Conclusion} \label{Section_Conclusion}

This paper proposes a general pre- and post-processing technique applicable
to any learning-based UL-to-DL CCM conversion method. A major bottleneck of
learning-based conversion is identified as the phase ramp produced by the mean
spatial frequency. By decomposing the problem into estimating the mean spatial
frequency and learning only the residual, we improve the CCM estimation
performance of three learning-based methods under uniform, Laplacian, and
Gaussian PAS models, as well as the resulting channel estimation performance.
The performance deterioration with an increasing number of base station
antennas for a constant dataset size is substantially reduced, and the
interpolation-based learners no longer fall behind the model-based benchmark
at large $M$. Future work includes multi-cluster channels and other array geometries,
such as the uniform rectangular array (URA), where the ramp becomes
two-dimensional, and the uniform circular array (UCA), where the
factorization must be reformulated.

\section*{Acknowledgment}
The author used Anthropic's Claude models as a research assistant.
The tool suggested the decomposition on which the proposed method is based,
drafted and revised parts of the simulation code, proposed the structure of Algorithm~\ref{alg:deramp}, drafted parts of
the text, and revised the language, consistency, and formatting of all
sections. All ideas were evaluated, corrected, and validated by the author
through independent simulations, and several intermediate claims produced by
the tool were found to be incorrect and discarded. The author designed the
experiments, verified all results, and takes full responsibility for the
content of this paper.

\bibliographystyle{IEEEtran}
\bibliography{references}

\end{document}